\documentclass[preprint,prd,aps,floats,nofootinbib,showpacs,floatfix]{revtex4-1}

\usepackage{graphicx}
\usepackage{epstopdf}
\usepackage[colorlinks=true, citecolor=blue, urlcolor = blue, linkcolor= blue, bookmarks=true]{hyperref}

\newcommand{\PRLett}[3]{Phys. Rev. Lett. {\bf #1}, #2 (#3).}
\newcommand{\PRD}[3]{Phys. Rev. D {\bf #1}, #2 (#3).}
\newcommand{\PRC}[3]{Phys. Rev. C {\bf #1}, #2 (#3).}
\newcommand{\PLB}[3]{Phys. Lett. B {\bf #1}, #2 (#3).}
\newcommand{\EPJA}[3]{Eur. Phys. J. A {\bf#1}, #2 (#3).}

\newcommand{\NPA}[3]{Nucl. Phys. A {\bf #1}, #2 (#3).}

\newcommand{\gam }{\ensuremath{\gamma }}
\newcommand{\gs }{\ensuremath{\sigma }}     
\newcommand{\gz }{\ensuremath{\zeta }}       
\newcommand{\gd }{\ensuremath{\delta }}      
\newcommand{\go }{\ensuremath{\omega }}   

\newcommand{\gn }{\ensuremath{\eta}}         

\newcommand{\gX }{\ensuremath{\Xi }}          
\newcommand{\gS }{\ensuremath{\Sigma }}   
\newcommand{\gL }{\ensuremath{\Lambda }} 

\newcommand{\fbs }{\ensuremath{f_{B_S}}}
\newcommand{\fds }{\ensuremath{f_{D_S}}}

\newcommand{\llags}[1]{\mathcal{L}_{#1}} 

\newcommand{\dmucon}{\ensuremath{\partial^\mu}} 
\newcommand{\dmucov}{\ensuremath{\partial_\mu}} 

\newcommand{\dsplus }{\ensuremath{D_S^+}}
\newcommand{\dsminus }{\ensuremath{D_S^-}}

\newcommand{\bszero }{\ensuremath{B_S^0}}
\newcommand{\bsbarzero }{\ensuremath{{\bar B}_S^0}}
\newcommand{\dsm }{\ensuremath{D_S} meson}
\newcommand{\dm }{\ensuremath{D} meson}

\newcommand{\bsm }{\ensuremath{B_S} meson}
\newcommand{\mds }{\ensuremath{m_{D_S}}}

\newcommand{\mbs }{\ensuremath{m_{B_S}}}

\newcommand{\lag }{Lagrangian}
\newcommand{\lagd }{Lagrangian density}

\newcommand{\wtt }{Weinberg-Tomozawa term}

\newcommand{\grz }{\ensuremath{\rho_0}}

\newcommand{\grb }{\ensuremath{\rho_B}}

\newcommand{\grsub}[1]{\ensuremath{\rho_{#1}}}
\newcommand{\grssub}[1]{\ensuremath{\rho_{#1}^s}}

\newcommand{\fs }{\ensuremath{f_s}}
\newcommand{\sfrac }{strangeness fraction}
\newcommand{\isoap }{isospin asymmetry parameter}

\newcommand{\chigrp}{\ensuremath{SU(3)_L\times SU(3)_R}} 

\begin{document}

%
\title{Bottom-strange mesons in hyperonic matter} 

\author{Divakar Pathak}
\email{dpdlin@gmail.com}
\affiliation{Department of Physics, Indian Institute of Technology, Delhi, 
Hauz Khas, New Delhi $-$ 110 016, India}

\author{Amruta Mishra}
\email{amruta@physics.iitd.ac.in}
\affiliation{Department of Physics, Indian Institute of Technology, Delhi,
Hauz Khas, New Delhi $-$ 110 016, India}

\begin{abstract}
The in-medium behavior of bottom-strange pseudoscalar mesons in hot, isospin asymmetric and dense hadronic environment is studied using a chiral effective model. 
The same was recently generalized to the heavy quark sector and employed to study the behavior of open-charm and open-bottom mesons. The heavy quark (anti-quark) is treated as frozen and all medium modifications of these bottom-strange mesons are due to their strange anti-quark (quark) content. We observe a pronounced dependence of their medium mass on baryonic density and strangeness content of the medium. Certain aspects of these in-medium interactions are similar to those observed for the strange-charmed mesons in a preceding investigation, such as the lifting of mass-degeneracy of $B_S^0$ and ${\bar B}_S^0$ mesons in hyperonic matter, while the same is respected in vacuum as well as in nuclear matter. In general, however, there is a remarkable distinction between the two species, even though the formalism predicts a completely analogous in-medium interaction Lagrangian density. We discuss in detail the reason for different in-medium behavior of these bottom-strange mesons as compared to charmed-strange mesons, despite the dynamics of the heavy quark 
being treated as frozen in both cases. 
\end{abstract}


\maketitle

\begin{section}{Introduction}\label{introduction}

The method of chiral invariant \lag s is a popular research strategy, often employed to understand the in-medium properties of 
hadrons. The general formulation for writing a \lag \ invariant under a particular Lie group goes back to the 
work of 
Coleman et al. \cite{coleman1, coleman2}. In the context of the study of hadronic in-medium interactions, the same ideology is utilized in a particular phenomenological effective \lag \ approach, wherein the Lie group in question is the chiral \chigrp \ \cite{Pap_prc99}. 
This has proved to be a fairly successful strategy, which has been extensively applied over the years to elucidate various aspects of strong interaction physics in this non-perturbative regime of QCD. 
Particularly, this is well suited to shed light on the behavior of matter under extreme conditions of density and temperature, which is relevant from the point of view of ongoing and future relativistic heavy ion collision experiments. 
Apart from a reasonable description of nuclear matter, finite nuclei, neutron stars and hypernuclei \cite{Zsch}, the model has been used to study the in-medium interactions of light vector mesons \cite{mamZsch_vector2004, mam_balazs_vector2004}, kaons and antikaons \cite{mamK2004, mam_kaons2006, mam_kaons2008, sambuddha1, sambuddha2}, as well as, on subsequent generalization, \dm s \cite{mamD2004, arindam, arvDprc, arvDepja}. Generalization of this framework to $SU(4)$ \cite{arvDepja, hofmannlutz} was an important step forward, since \dm s are intrinsically different from, e.g. kaons and antikaons, due to the presence of a heavy charm quark/antiquark. Thus, while the framework was generalized to $SU(4)$, the medium modifications of such heavy-light mesons were completely borne out of the light quark content, whereas the heavy quark degrees of freedom were treated as frozen in the medium. This generalization led to progress on two counts -- on one hand, we were equipped with a formalism to treat medium effects for strange-charmed mesons \cite{DP_Dsm_arxiv} which were hitherto unconsidered. Secondly, it inspired a further generalization to bottomed mesons \cite{DP_Bm_arxiv}, based on the general philosophy that if the dynamics of the heavy quark is frozen, it is immaterial whether that heavy quark is $c$ or $b$. Thus, just like their charmed counterparts, bottomed mesons can also be treated as a heavy-light system, where all medium effects are due to interaction of their light quark content with the light hadrons in the medium. 
Clearly then, the final class of pseudoscalar mesons that can be treated within this generalized formalism, are the \emph{bottom-strange} ($B_S$) mesons. The current article is an attempt to unveil their in-medium behavior.  

We organize this article as follows: in section II, we give a brief outline of the chiral effective approach and write down the \lagd \ as well as the in-medium dispersion relations for the \bsm s, within this model. In section III, we analyze and discuss the in-medium behavior of \bsm s as per this formulation. Finally, the entire investigation is summarized in section IV. 
\end{section}

\begin{section}{\protect \boldmath \texorpdfstring{$B_S$}{Bs} Mesons in Hadronic Matter -- Chiral Effective Approach}

As mentioned previously, the present work is based on a generalization of the chiral \chigrp \ model \cite{Pap_prc99, Zsch}. This is an effective field theoretical approach to describe hadron-hadron interactions, based on the approximate chiral invariance of the strong interactions. The fact that chiral symmetry is explicitly broken is taken care of, with the addition of explicit symmetry breaking terms to the chiral-symmetric \lag. Scale symmetry breaking is introduced via a scalar dilaton field, $\chi$, leading to a non-vanishing trace of the energy-momentum tensor \cite{Zsch, DP_Bmonia}. Also, following Weinberg \cite{weinberg67, weinberg68}, a nonlinear realization of chiral symmetry \cite{bardeenlee} is adopted in this effective hadronic description. 

The general expression for the chiral model \lagd \ reads: 
\begin{equation}
{\cal L} = {\cal L}_{\rm kin}+\sum_{\rm W} {\cal L}_{\rm BW} + {\cal L}_{\rm vec} + {\cal L}_{\rm 0} + {\cal L}_{\rm scale break} + {\cal L}_{\rm SB}
\label{genlag_model}
\end{equation}
In Eq.(\ref{genlag_model}), $\llags{\rm kin}$ is the kinetic energy term. $\llags{\rm BW}$ represents the interaction of baryons with mesons of the type $W$, where $W$ covers both spin-0 and spin-1 mesons, the former generating the baryon masses. $\llags{\rm vec}$ represents the interaction term for scalar mesons with vector mesons, which dynamically generates the vector meson masses, as well as the self-interaction terms for these mesons. $\llags{\rm 0}$ induces spontaneous breaking of chiral symmetry via meson-meson interactions. $\llags{\rm scalebreak}$ introduces scale invariance breaking, via a logarithmic potential 
term in the scalar dilaton field, $\chi$, as was mentioned above. Finally, $\llags{\rm SB}$ is the explicit symmetry breaking term. 

From this chiral model \lagd, in the mean field approximation, one determines the equations of motion for the non-strange scalar-isoscalar meson $\gs (\sim {\bar u}u + {\bar d}d)$, strange scalar-isoscalar meson $\gz (\sim {\bar s}s)$, the non-strange scalar-isovector meson $\gd (\sim {\bar u}u - {\bar d}d)$, and the scalar dilaton 
$\chi$ \cite{Pap_prc99, Zsch, arvDprc}. 
These coupled equations of motion are solved self-consistently and the 
solutions so obtained are used to evaluate the baryonic scalar and 
number densities. These are subsequently employed in the solution 
of the in-medium dispersion relations for the \bsm s, 
which we address shortly. 
%
%
As mentioned in section I, this approach, which was originally devised for chiral \chigrp, has subsequently been generalized to include charm \cite{arvDepja, hofmannlutz} and bottom \cite{DP_Bm_arxiv} degrees of freedom as well. The interaction \lagd \ for pseudoscalar mesons has contributions from a vectorial \wtt, scalar meson exchange contributions originating from the explicit symmetry breaking term, as well as range terms. Each of these 
was 
elaborately discussed in a preceding work \cite{DP_Bm_arxiv} and hence, 
for the sake of brevity, will not be repeated here. Instead, we build on 
that formulation and proceed to discuss what it entails for the behavior 
of \bsm s in a hadronic medium. 
 
The \lagd \ for the \bsm s in isospin-asymmetric, strange, hadronic medium is given by the equation
\begin{equation}
\llags{\rm total} = \llags{\rm free} + \llags{\rm int},
\label{totalL} 
\end{equation}
where $\llags{\rm free}$ denotes the free \lagd \ for a complex scalar field (corresponding to the \bsm s here), which reads: 
\begin{equation}
\llags{\rm free} = \left( \dmucon \bsbarzero\right) \left(\dmucov \bszero\right) \ -\mbs^2 \left( \bsbarzero \bszero \right) 
\end{equation}
The interaction \lagd, $\llags{\rm int}$, is determined to be:
\begin{eqnarray}
\llags{\rm int} & = & -\frac {i}{4\fbs^2}\Big[\Big( 2\left( \bar \gX^0 \gam^{\mu} \gX^0 + \bar \gX^- \gam^{\mu}\gX^- \right) + \bar{\gL}^{0}\gam^{\mu}\gL^{0} + \bar{\gS}^{+}\gam^{\mu}\gS^{+} \nonumber\\ 
& & \ \ \ \ \ \ \ + \ \bar{\gS}^{0}\gam^{\mu}\gS^{0} + \bar{\gS}^- \gam^{\mu} \gS^- \Big) \Big(\bsbarzero (\dmucov \bszero) - (\dmucov \bsbarzero) \bszero \Big) \Big] \nonumber\\ 
& & + \frac{\mbs^2}{\sqrt2 \fbs} \Big[(\gz' +\gz'_b) \left( \bsbarzero \bszero \right) \Big] \nonumber\\ 
& & - \frac {\sqrt2}{\fbs}\Big[(\gz' +\gz'_b)\ \Big( (\dmucov \bsbarzero)(\dmucon \bszero)\Big)\Big] \nonumber \\
& & + \frac {d_1}{2 \fbs^2}\Big[\Big( \bar p p +\bar n n +\bar{\gL}^{0}\gL^{0}+\bar{\gS}^{+}\gS^{+}+\bar{\gS}^{0}\gS^{0}\nonumber\\
& & \ \ \ \ \ \ \ \ \ \ \ \ \ \ + \ \bar{\gS}^{-}\gS^{-} + \bar{\gX}^{0}\gX^{0}+\bar{\gX}^{-}\gX^{-}\Big)  \Big( (\dmucov \bsbarzero)(\dmucon \bszero) \Big)\Big]\nonumber \\
& & + \frac {d_2}{2 \fbs^2} \Big [\Big( 2\left( \bar \gX^0 \gX^0 + \bar \gX^- \gX^- \right) + \bar{\gL}^{0}\gL^{0}+\bar{\gS}^{+}\gS^{+} \nonumber\\ 
& & \ \ \ \ \ \ \ \ \ \ \ \ \ \ \ \ \ \ \ \ \ +\  \bar{\gS}^{0}\gS^{0} + \bar{\gS}^- \gS^-\Big)\Big((\dmucov \bsbarzero)(\dmucon \bszero)\Big)\Big] 
\label{L_int_bs}
\end{eqnarray}
In this expression, the first term (with coefficient $-i/4\fbs^2$) 
is the \wtt, obtained from the kinetic energy term in 
Eq. (\ref{genlag_model}), the second term (with coefficient 
$\mbs^2/\sqrt{2}\fbs$) is the scalar meson exchange term, obtained 
from the explicit symmetry breaking term of the \lag, third term 
(with coefficient $-\sqrt{2}/\fbs$) is the first range term, which 
is obtained from the kinetic energy term of the pseudoscalar mesons. 
This, along with the fourth and fifth terms (the $d_1$ and $d_2$ range 
terms \cite{DP_Bm_arxiv}), represents the total range term in the chiral 
model. Also, $\gz' \ (= \gz - \gz_0)$, and $\gz_b' \ (= \gz_b - \gz_{b0})$
 denote the fluctuations of the strange and bottomed scalar fields 
from their respective vacuum values. 

Using the Euler-Lagrange equation on Eq.(\ref{totalL}), we arrive at the equations of motion for the \bszero \ and \bsbarzero \ mesons. It can be straightaway inferred from Eq.(\ref{L_int_bs}) that these equations of motion are linear in these fields. Hence, exploiting this linearity, we assume plane wave solutions $( \sim \ e^{i(\vec{k}.\vec{r} - \go t)} )$ and Fourier transform these equations to obtain the in-medium dispersion relations for the \bsm s.
These have the general form:
\begin{equation}
-\go^{2} + \vec{k}^2 + \mbs^2 - \Pi (\go,|\vec{k}| ) = 0
\label{dispersion}
\end{equation}
where \mbs \ refers to the \bsm \ vacuum mass, and $\Pi (\go,| \vec{k} |)$ refers to the 
\emph{self-energy} of the \bsm s in the medium. A complete expression for the \bsm \ self-energy is given by:
\begin{eqnarray}
\Pi (\go, |\vec k|) &=& \Bigg[ \Bigg( \frac{d_1}{2\fbs^2} \big( \grssub{p} + \grssub{n} + \grssub{\gL} + \grssub{\gS^+} + \grssub{\gS^0} + \grssub{\gS^-} + \grssub{\gX^0} + \grssub{\gX^-}\big) \Bigg) \nonumber\\
 & & \ \ + \Bigg( \frac{d_2}{2\fbs^2} \big( 2(\grssub{\gX^0} + \grssub{\gX^-}) + \grssub{\gL} + \grssub{\gS^+} + \grssub{\gS^0} + \grssub{\gS^-}\big) \Bigg) \nonumber \\
& & \ \ -\Bigg( \frac{\sqrt{2}}{\fbs}\big( \gz' + \gz_b' \big) \Bigg) \Bigg]\Big( \go ^2 - {\vec k}^2 \Big) \nonumber\\
& & \pm \Bigg[ \frac{1}{2\fbs^2} \Big( 2(\grsub{\gX^0} + \grsub{\gX^-}) + \grsub{\gL} + \grsub{\gS^+} + \grsub{\gS^0} + \grsub{\gS^-} \Big) \Bigg] \go \nonumber\\
& & + \Bigg[ \frac{\mbs^2}{\sqrt{2}\fbs}\big( \gz' + \gz_b' \big) \Bigg]
\label{selfenergy}
\end{eqnarray}
where the $+$ and $-$ signs in the coefficient of the linear term, refer to \bsbarzero \ and \bszero \  respectively. Thus, it is explicit from this equation that this term is all that differentiates between the two, the remaining interaction terms being identical for both. We point out that the source of this contribution can be traced back to the \wtt \ in Eq.(\ref{L_int_bs}). This leads to very interesting consequences that shall be explored in the next section. Proceeding further, 
these dispersion relations reduce to the following quadratic equation in the centre-of-mass frame of these mesons:
\begin{equation}
-\go^2 + \mbs^2 - \Pi \left( \go, 0 \right) = 0,
\label{disp_relatn_k0_quad}
\end{equation}
This quadratic equation is solved to find out the medium modified mass (energy) $\go(\vec k = 0)$ for the \bsm s. Also, in line with our prior research 
methodology \cite{DP_Dsm_arxiv, DP_Bm_arxiv}, we study the sensitivity of these medium modifications to the following four parameters - the total baryonic density (\grb) of the medium, temperature, \isoap \ defined as $\gn \ = -\sum_{i}  I_{3i} \rho_{i}/\grb$,
where $I_{3i}$ refers to the $z-$component of isospin of the $ i^{th}$ baryon,
and, \sfrac, $f_s = \sum_{i}  |S_{i}| \rho_{i}/\rho_{B}$, where $S_{i}$ denotes the strangeness quantum number of the $i^{th}$ baryon.

We point out here that the above expressions for the \bsm \ interaction \lagd, as well as in-medium self-energy, are exactly analogous to those for the \dsm s \cite{DP_Dsm_arxiv}. One feels tempted to invoke the similarity between their quark content at this stage, and expect the complete in-medium behavior to follow suite. However, in spite of this analogy, there exist differences between them, as we describe, along with other aspects of the in-medium behavior, in the next section.  
\end{section}

\begin{section}{Results and Discussion} \label{results_section}
Before proceeding with an account of the in-medium properties of \bsm s, we first make a few general remarks, especially concerning our parameter choice. The parameters of the chiral model are fitted to vacuum baryon masses, nuclear saturation properties and other vacuum characteristics in the mean field limit \cite{Zsch, Pap_prc99}. We maintain here the same parameter choice that was previously used to study charmed \cite{arindam, arvDprc, arvDepja}, charmed-strange \cite{DP_Dsm_arxiv} and open-bottom mesons \cite{DP_Bm_arxiv}, and refer the interested reader to 
earlier works \cite{arvDepja, Pap_prc99} for a summary of these fitting procedures. 
Also, from equations (\ref{L_int_bs}) and (\ref{selfenergy}), one can reason that the \bsm \ decay constant is a crucial parameter in this analysis, since that would directly impact the magnitude of these medium effects. In the absence of a Particle Data Group \cite{PDG2012, PDG_website} value for the same, in this investigation, we employ $\fbs = 224$ MeV, consistent with Ref.\cite{fbs224_PRL2013}. Similar 
values 
were also obtained in other recent lattice results \cite{fbs_HPQCD_PRD2012, fbs_ETM_JHEP2014, fbs_alpha_NPB2013, fbs_ETM_JHEP2012}, as well as in recent calculations based on the QCD Sum Rule method \cite{fbs_Lucha_PRD2013, fbs_Lucha_JPhysG2011, fbs_Narison_PLB2013, fbs_Khodjamirian_PRD2013}. 
The only simplifying approximation adopted in this work is the neglect 
of the heavy quark dynamics. This approximation, which appears justified 
even on physical grounds, has also been adopted in each and every work 
\cite{arindam, arvDprc, arvDepja, DP_Dsm_arxiv, DP_Bm_arxiv} concerning 
the generalization of this chiral $SU(3)$ approach to heavy quark sector 
($c$ and $b$). In the context of charmed quark condensate, the same was 
also explicitly verified in a O(4) linear sigma model \cite{roder}. 
Also, we 
point out that the \emph{`frozen-glueball approximation'} 
$(\chi \approx \chi_0)$ \cite{Zsch}, which was adopted in our earlier
 work \cite{DP_Dsm_arxiv} concerning the \dsm s, stands relaxed in this 
investigation. 
Thus, the scalar dilaton field ($\chi$) is also considered to be medium-modified in this work.   

We now discuss the \bsm \ in-medium properties as per the preceding 
formalism. It follows from Eq.(\ref{selfenergy}), that just like their 
charmed counterparts \cite{DP_Dsm_arxiv}, even the two \bsm  s, 
\bszero \ and \bsbarzero, are degenerate in nuclear matter. 
This is because the only term in the interaction \lagd \ that distinguishes between these two is the vectorial \wtt, whose magnitude depends exclusively on hyperonic contributions. The same is also true for the attractive $d_2$ range term, though it contributes equally to them.  
However, one expects this mass degeneracy to get lifted upon the addition of hyperons to the medium, with this Weinberg-Tomozawa producing an equal increase/decrease in the medium mass for \bszero \ and \bsbarzero \ mesons. This behavior is reflected in Fig. \ref{Bs_TbyT} wherein the various individual contributions to the total in-medium mass of the \bsm s is shown, for hyperonic matter ($\fs  = 0.5$) at different temperatures, as a function of baryonic density \grb. It can be clearly seen that the disparity between \bszero \ and \bsbarzero \ originates through this Weinberg-Tomozawa contribution only while all other interaction terms contribute equally to them. Overall, the medium mass of both \bsm s is observed to decrease, though asymmetrically, owing to the aforementioned disparity. For example, in asymmetric ($\eta = 0.5$) hyperonic matter, with $\fs = 0.5$, at $T=0$, \bszero \ meson is observed to experience a mass drop of $67$, $161$ and $326$ MeV at $\grb = \grz$, $2\grz$ and $4\grz$ respectively, while the corresponding numbers for \bsbarzero \ read  $73$, $173$ and $349$ MeV. Thus, there is a significant density dependence of the in-medium mass, which is anyhow expected from the \bsm \ in-medium dispersion relations, equations (\ref{dispersion}) and (\ref{selfenergy}). A reduction in the net in-medium mass signifies a net attractive in-medium interaction, though the behavior of individual contributions is not that mundane. 
In fact, we may reason from Eq.(\ref{selfenergy}) and the in-medium behavior of scalar fields \cite{arvDprc}, that the first range term contribution is repulsive throughout, while those of the other two range terms and the scalar meson exchange term is attractive for the entire range of baryonic densities considered in this analysis. These attractive contributions from the $d_1$ and $d_2$ range terms predominate over the repulsive first range term contribution, and we observe the net behavior as shown in the figure. However, to provide a more concrete basis to these assertions, as well as to weigh the relative significance of these individual contributions towards the total in-medium mass, we analyze the \bsm \ in-medium dispersion relations as follows. 

\begin{figure}
\begin{center}
\scalebox{0.75}
{\includegraphics{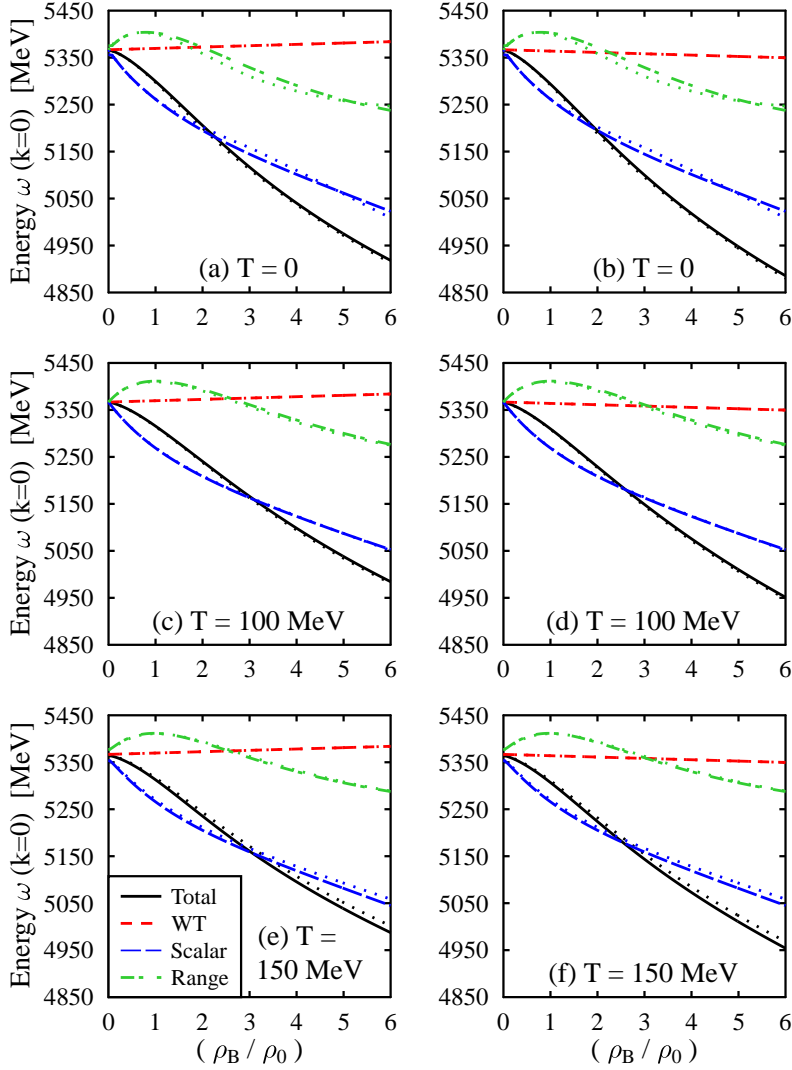}}\caption{ \label{Bs_TbyT} (Color Online) The various contributions to the energy at ${\vec k} = 0$, for the \bsm s in hyperonic matter ($f_s = 0.5$), at different temperatures. Subplots (a), (c) and (e) correspond to the 
\bszero \ meson, while (b), (d) and (f) correspond to the 
\bsbarzero \ meson. For each case, the variation of the individual contributions in asymmetric matter (with $\eta = 0.5$), as described in the legend, is also compared against the symmetric situation ($\eta = 0$), shown as dotted lines.}
\end{center}
\end{figure}

As mentioned previously, in the center of mass frame of these mesons, the in-medium dispersion relations acquire the quadratic equation form $A\go^2 + B\go + C = 0$, where the coefficients $A, \ B$ and $C$ depend on various interaction terms in Eq.(\ref{dispersion}), and read: 
\begin{eqnarray} 
A & = & \Bigg[ 1 + \Big( \frac{d_1}{2\fbs^2} \sum_{(N+H)}\grssub{i} \Big) + \Bigg( \frac{d_2}{2\fbs^2} \Big( 2 \sum_{\gX}\grssub{i} \ + \sum_{\gS, \gL}\grssub{i} \Big) \Bigg) \nonumber\\ 
& & \hspace{3cm} -\Big( \frac{\sqrt{2}}{\fbs}\big( \gz' + \gz_b' \big) \Big) \Bigg] \ \ \ 
\label{disp_a}
\end{eqnarray}
\begin{equation}
B =  \pm \Bigg[ \frac{1}{2\fbs^2} \Big(  2 \sum_{\gX}\grsub{i} \ + \sum_{\gS, \gL}\grsub{i} \Big) \Bigg] 
\label{disp_b}
\end{equation}
\begin{equation}
C = \Bigg[-\mbs^2 +  \frac{\mbs^2}{\sqrt{2}\fbs}\big( \gz' + \gz_b' \big) \Bigg],
\label{disp_c}
\end{equation}
where the $+$ and $-$ signs in the coefficient $B$, correspond to \bsbarzero \ and \bszero, respectively. The above summations are to be read as follows: in the first term, $(N+H)$ denotes the sum over all baryons, while 
$\gX$ denotes a sum over the two Xi hyperons  $(\gX^{-,0})$. Likewise, $(\gS, \gL)$ is to be read as a sum over the set $(\gS^{\pm,0}, \gL)$. 
Just like the \dsm \ case \cite{DP_Dsm_arxiv}, 
we write the general solution of this quadratic equation, as:
\begin{eqnarray}
\go = \frac{(-B + \sqrt{B^2 - 4AC})}{2A} \approx  -\frac{B}{2A} 
+ {\sqrt {\frac{C_1}{A}}} + \frac{B^2}{8A\sqrt{AC_1}} + \ldots
\end{eqnarray}
using a binomial expansion of $(1 + B^2/4AC_1)^{1/2}$, with $C_1 = -C$. Disregarding higher-order terms, this can be simply written as:
\begin{equation}
\go \approx {\sqrt {\frac{C_1}{A}}}  -\frac{B}{2A} 
\label{sign_reversal_eqn}
\end{equation}
With similar expansions of $(C_1/A)^{1/2}$ and $(1/A)$, and neglecting higher order contributions (which are smaller, owing to inverse dependence on large powers of \fbs), 
the general solution of the \bsm \ dispersion relations in the hyperonic matter context, can be written as: 
\begin{equation}
\go = \go_{\rm com} \mp \go_{\rm brk},
\end{equation} 
where 
\begin{eqnarray}
\go_{\rm com} & = & \sqrt{\frac{C_1}{A}} \approx \mbs \left[1 - \left( \frac{d_1}{4\fbs^2} \sum_{(N+H)}\grssub{i} \right) \right.\nonumber\\ 
& & - \left. \left( \frac{d_2}{4\fbs^2} \left( 2 \sum_{\Xi}\grssub{i} + \sum_{\gS,\gL}\grssub{i} \right) \right) +\left( \frac{\left( \gz' + \gz_b' \right)}{2\sqrt{2}\fbs} \right) \right],
\label{commonpartofthesolution}
\end{eqnarray}
\begin{equation}
\go_{\rm brk} = \frac{|B|}{2A} \approx \frac{|B|}{2} = \Bigg[ \frac{1}{4\fbs^2} \Big(  2 \sum_{\Xi}\grsub{i} \ + \sum_{\gS,\gL}\grsub{i} \Big) \Bigg]
\label{breakingpartofthesolution}
\end{equation}
The former contributes equally to both \bszero \ and \bsbarzero, while the latter, which can be immediately recognized as the contribution emanating from the \wtt \ only, gives equal and opposite contributions to the two. Thus, $\go_{\bszero} = \go_{\rm com} + \go_{\rm brk}$, and $\go_{\bsbarzero} = \go_{\rm com} - \go_{\rm brk}$, (where $\go_{\rm brk} \geq 0$), these extra contributions being responsible for the larger mass drop of \bsbarzero \ meson as compared to \bszero, as observed in Fig. \ref{Bs_TbyT}. 

Some aspects of this observed in-medium behavior of the \bsm s are 
strikingly similar to those for the \dsm s, which bodes well with 
the analogy we had drawn earlier. For example, a weak dependence 
of the medium mass on isospin asymmetry is apparent from 
Fig. \ref{Bs_TbyT}. This looks consistent with the apparent isospin 
symmetry of the self-energies. However, as was elaborately discussed 
in the context of the \dsm s \cite{DP_Dsm_arxiv}, the fact that the medium mass depends at all on isospin asymmetry is because the equation of motion for the strange condensate, obtained within this chiral model within the mean field approximation, is coupled to the scalar-isovector $\gd$ field \cite{arvDprc}, and hence, acquires different values for isospin symmetric and isospin asymmetric matter. But since this happens to be the source of asymmetry here, and not any explicitly isospin-asymmetric term in the \bsm \ self-energies, it is natural that this asymmetry would be weak, which is absolutely consistent with Fig. \ref{Bs_TbyT}. 
Likewise, a reduction in the magnitude of the scalar fields' fluctuations from their vacuum values with temperature, ensures a reduction in the magnitude of the mass drops with increasing temperature. For example, we observe from Fig. \ref{Bs_TbyT} that the mass drops of \bszero \ and \bsbarzero \ mesons, in asymmetric ($\eta = 0.5$), hyperonic ($f_s = 0.5$) matter, at $\grb = 6\grz$ and $T=0$ read $448$ and $481$ MeV respectively, which reduce to $379$ and $413$ MeV respectively when the temperature increases to $T=150$ MeV. This is not just consistent with the above reasoning, but also with the general conclusion that medium effects decrease in magnitude with temperature, as has been observed in several investigations based on this chiral effective approach \cite{mamK2004, mamD2004, arvDprc, arvDepja, DP_Dsm_arxiv, DP_Bm_arxiv, DP_Bmonia}. 

\begin{figure}
\begin{center}
\scalebox{0.75}
{\includegraphics{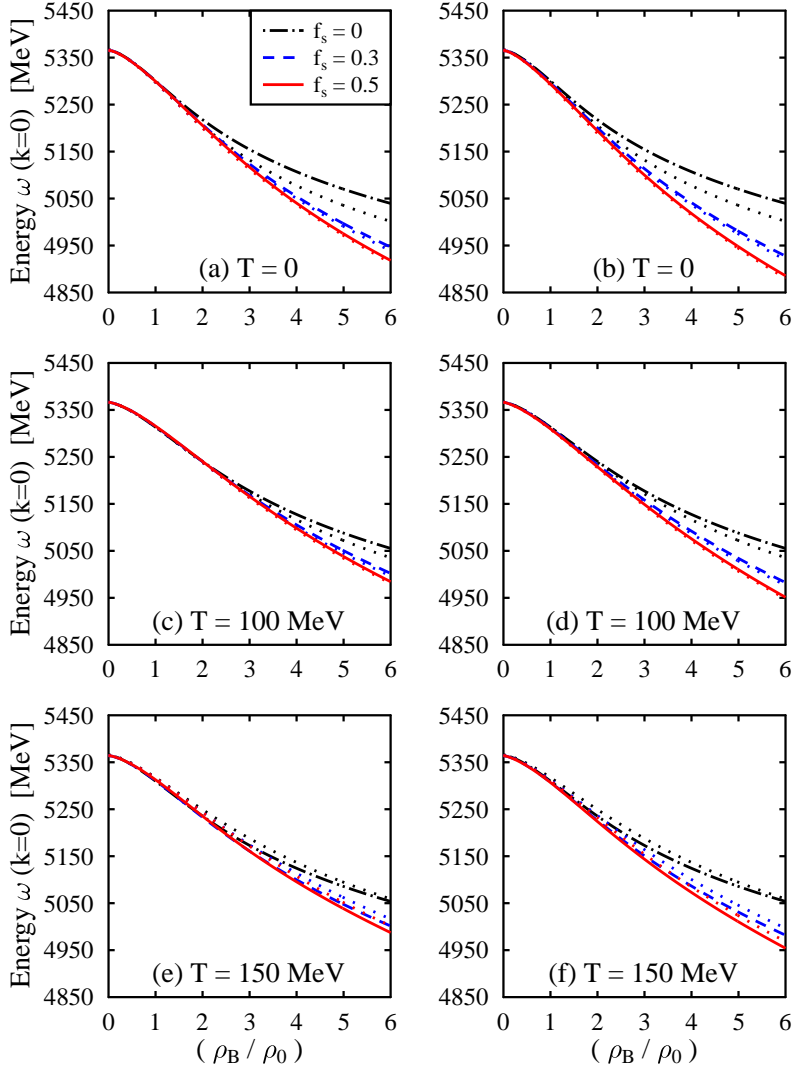}}\caption{ \label{Bs_fsvary} (Color Online) A comparison of the energy at ${\vec k} = 0$, of the \bszero \ (subplots (a), (c) and (e)) and \bsbarzero \ meson (subplots (b), (d) and (f)), for different values of $f_s$, and at different temperatures. In each case, the isospin asymmetric situation ($\eta = 0.5$), as described in the legend, is also compared against the symmetric ($\eta = 0$) situation, represented by dotted lines.}
\end{center}
\end{figure}

However, there is a considerable dependence of the \bsm \ in-medium mass on the strangeness content of the medium, as is analyzed in Fig. \ref{Bs_fsvary}, for both symmetric ($\eta = 0$) and asymmetric ($\eta = 0.5$) matter, at different temperatures. These plots also subsume the special case of nuclear matter ($f_s = 0$), where \bszero \ and \bsbarzero \ are identical due to the vanishing of $\go_{\rm brk}$. We observe from the figure that there is a considerable reduction in the medium mass for both, as the value of \fs \ is increased. For example, the medium masses of \bszero \ and \bsbarzero \ in cold $(T=0)$ asymmetric $(\eta = 0.5)$ hyperonic matter with $f_s = 0.5$, at $\grb = 6\grz$, are $4918$ and $4886$ MeV, as compared to their nuclear matter value of $5040$ MeV. This behavior is because of the extra hyperonic contributions to $d_1$ and $d_2$ range terms, which make them more attractive in the hyperonic matter context. Also, the effect of \wtt \ (or $\go_{\rm brk}$, from the point of view of the closed form solution) giving a larger drop to \bsbarzero \ as compared to \bszero \ is clearly reflected in Fig. \ref{Bs_fsvary}. 

Next, we address the disparity between the \emph{strange-charmed} \dsm s, and the \emph{strange-bottomed} \bsm s, with regard to their observed in-medium behavior. As we have been insisting all along, we intuitively expect their in-medium behavior to be similar to one another, since in each case, the heavy quark is frozen in the medium and all modifications are due to the light quark content, which is identical for them. Further, the interaction \lagd \ for \bszero \ and \bsbarzero \ is found to be completely analogous to that for \dsminus \ and \dsplus \ respectively, as was mentioned in the previous section. This analogy notwithstanding, there is a conspicuous difference in the observed medium-behavior in the two cases. For example, for the \dsm s, it was observed that the \fs \ dependence of \dsplus \ was highly pronounced, while that of \dsminus \ was almost negligible, due to a near-cancellation of the respective \fs \ dependences of $\go_{\rm com}$ and $\go_{\rm brk}$ \cite{DP_Dsm_arxiv}. However, over here, we observe that the \fs \ dependence of both \bsm s is appreciable and no such cancellation takes place. This is due to the fact that the magnitude of the  $\go_{\rm com}$ contribution is much larger than that of $\go_{\rm brk}$. 
To further appreciate this observed behavior, we consider the transition from \dsm s to \bsm s from the point of view of various interaction terms in the \lagd. With this transition, the magnitude of the \wtt, as well as the $d_1$ and $d_2$ range terms, changes by a factor of $(\fds/\fbs)^2$, while that of the first range term changes by $(\fds/\fbs)$. With the adopted values of parameters here, as well as in Ref.\cite{DP_Dsm_arxiv}, these factors are both close to unity, being $1.1$ and $1.05$ respectively. On the other hand, the scalar meson exchange contribution would change by a factor of $\left( (\mbs/\mds)^2 \times (\fds/\fbs) \right) \approx 2.86$, which is appreciably larger. Thus, at fixed values of the parameters $(\grb, T, \eta, \fs)$, while the contributions from the range terms and the \wtt \ for the \dsm s and \bsm s are comparable, the attractive scalar meson exchange contribution is significantly larger for the latter. On one hand, this is responsible for the much bigger mass drops of the latter as compared to the former. 
Secondly, it may be noticed that this scalar meson exchange term contributes completely to $\go_{\rm com}$, while $\go_{\rm brk}$ depends only on the Weinberg-Tomozawa contribution which has approximately the same magnitude for both \dsm s and \bsm s. This implies that while the mass asymmetry between the two \bsm s continues to be of the same order as that for the \dsm s, the mass drops are significantly intensified for the former. 
For example, 
we had observed mass drops of $161$ and $131$ MeV from the vacuum value for the \dsplus \ and \dsminus \ mesons respectively, in asymmetric ($\eta = 0.5$) hyperonic matter ($\fs = 0.5$) at $\grb = 6\grz$ and $T=100$ MeV \cite{DP_Dsm_arxiv}. Corresponding values for \bszero \ and \bsbarzero \ mesons in this investigation are observed to be $383$ and $416$ MeV respectively, which bears testimony to what was qualitatively deduced above. In fact, with the aforementioned reasoning, the scalar meson exchange term assumes a larger significance in this case, as compared to the \dsm s. Thus, we can draw the general conclusion that the interplay of various individual contributions to the total in-medium mass is different for the two cases, which leads to a conspicuous difference in the net observed behavior. 
In hindsight, the same appears natural, since with the replacement $c \leftrightarrow b$, one should only expect the in-medium behavior to be similar if the interaction \lagd \ scales by some overall constant factor with the meson (vacuum) mass and decay constant, which is clearly not the case here, as can be seen from Eq.(\ref{L_int_bs}). If that was the case, the interplay of all individual contributions would have also been unaffected, and we would have observed the same in-medium behavior for $D_S$ and \bsm s, perhaps scaled up or down by that constant factor. But since the individual contributions scale unequally, it is inevitable that their in-medium behavior would be dissimilar.   

We have, thus, analyzed the in-medium behavior of bottom strange mesons, as a function of various characteristic parameters. We find an appreciable dependence of their medium mass on baryonic density and strangeness content of the medium. These medium effects can potentially be important in the propagation and flow of these bottom-strange mesons in heavy-ion collision experiments, besides affecting observables like particle ratios and dilepton spectra, owing to the semileptonic/ leptonic decay modes of \bsm s.   
\end{section}

\begin{section}{Summary} 
To summarize, we have studied the in-medium properties of \bsm s in a hot and dense hadronic environment. Towards this end, we have considered a generalization of the chiral effective approach to the heavy quark sector. The heavy quark (anti-quark) is treated as frozen and all medium modifications of these bottom-strange mesons are due to their strange anti-quark (quark) content. The interaction \lagd, as well as the in-medium dispersion relations bear a striking resemblance with those obtained for the strange-charmed mesons in a preceding investigation \cite{DP_Dsm_arxiv}. However, in spite of this analogy, the behavior of \bsm s differs in some regard from that of the \dsm s, especially with the pronounced \fs \ dependence for both \bszero \ and \bsbarzero \ mesons. Certain other features, such as their mass degeneracy in vacuum, as well as in the nuclear matter situation, getting lifted upon the addition of hyperons, is exactly on the same lines as their charmed counterparts. Overall, we have studied the sensitivity of these medium modifications to baryonic density, strangeness content of the medium, isospin asymmetry and temperature. The former two are observed to significantly affect the observed in-medium behavior. While there has been some work \cite{DP_Bmonia, DP_Bm_arxiv, Hilger_QSR_D_Bm, Bm_QMC_Tsushima, YasuiSudoh_Bm_2014, YasuiSudoh_Bm_2013_1, YasuiSudoh_Bm_2013_2, Tolos_Bm} aimed at studying the behavior of hidden and open-bottom mesons, to the best of our knowledge, this is first time that the in-medium properties of bottom-strange pseudoscalar mesons have been investigated as a function of the aforementioned parameters. 
\end{section}  

\begin{section}*{Acknowledgments} 
D.P. acknowledges financial support from University Grants Commission, India [Sr. No. 2121051124, Ref. No. 19-12/2010(i)EU-IV]. A.M. would like to thank Department of Science and Technology, Government of India (Project No. SR/S2/HEP-031/2010) for financial support. 

\end{section}


\end{document}